\documentstyle[12pt,a4]{article}

\begin{document}

\title{\sc THE QUANTUM MECHANICAL FOUNDATIONS OF PHILOSOPHY }
\author{Cihan Sa\c{c}l\i o\~{g}lu$^{1,2}$\\
\date{$^{1}$Physics Department, Bo\~{g}azi\c{c}i University \\
80815 Bebek--\.{I}stanbul, Turkey\\
and \\
$^{2}$Feza G\"{u}rsey Institute \\   Bo\~{g}azi\c{c}i
University--TUBITAK  \\ 81220 \c{C}engelk\"{o}y, {I}stanbul--
Turkey}} \maketitle
%\vspace*{1cm}
\begin{abstract}

Many of the most familiar features of our everyday environment,
and some of our basic notions about it, stem from Relativistic
Quantum Field Theory (RQFT). We argue in particular that the
origin of common names, verbs, adjectives such as full and empty,
the concepts of identity, similarity, Plato's Universals, natural
numbers, and existence versus non-existence can be traced to the
space-time and gauge symmetries and quantum properties embodied in
RQFT. These basic tools of human thought cannot arise in a
universe strictly described by classical Physics based on Planck's
constant being exactly equal to zero.

\end{abstract}
\vspace*{3 cm} \pagebreak \baselineskip=30pt

\noindent  {\bf 1.  Introduction:}

A more accurate version of the title, requiring the sacrifice of
the obviously intended reference to some well-known titles
\cite{REF1}, would have been ``The Relativistic Quantum Field
Theoretical Foundations of Philosophy". To put it briefly, the
majority of works in which the words ``Philosophy" and ``Quantum
Mechanics" appear simultaneously focus on the aspects of Quantum
Mechanics that seem to be in conflict with common sense and
everyday (i.e., non-laboratory) experience, whereas the aim here
is to argue the opposite proposition that our everyday experiences
and common sense are shaped by Relativistic Quantum Field Theory
in a quite direct way.

Although the more familiar subject, which often goes under titles
such as ``The Philosophical Foundations of Quantum Mechanics", is
thus unrelated to the issues we intend to address here, it is
perhaps worthwhile to start with a few remarks about it in order
to delineate more clearly the domain of our argument. At the risk
of considerable simplification, one may say that the basic issue
in works along the lines of \cite{REF1} is the reconciliation of
our usual idea of ``existing objective reality" with the Quantum
Mechanical picture of different ``potential realities" that
simultaneously coexist prior to the observation-induced collapse
of the wave function. The intellectual climate among physicists
has only relatively recently become tolerant towards suggestions
that the Copenhagen Credo (in terms of which the preceding
sentence has been formulated) could perhaps be modified or
improved upon, and this has led to significantly more detailed
understanding of how the passage from a Quantum Mechanical
probability amplitude to classical, objective probability occurs
through decoherence \cite{REF2}. In short, some 70 years after the
formulation of the Copenhagen interpretation, the examination of
the philosophical concepts underlying Quantum Mechanics (QM)
continues, but the ``paradoxes" typically arise only in situations
where a coherent sum of states or amplitudes is involved.

Returning to, and somewhat rephrasing, the first paragraph, the
aim of the present note is to investigate the roughly reverse
question of to what extent some of our most basic everyday
notions, and the philosophical and mathematical concepts that are
abstracted from them, owe their existence to QM, or more
precisely, to a body of physical facts that find their most
natural mathematical expression in Relativistic Quantum Field
Theory (RQFT). These physical facts involve none of the
controversial issues addressed in \cite{REF1} and \cite{REF2}, or
the difficulties connected with the renormalization procedure in
RQFT. Instead, they are based on the most fundamental and commonly
accepted aspects of RQFT.  The reason we are not in the habit of
relating our everyday experiences to RQFT is twofold: (a) The very
familiarity of these experiences blunts our curiosity;
furthermore, as is often the case, this familiarity is confused
with our supposed ability to describe everyday macroscopic
phenomena in terms of classical physics. (b) Most physicists (and
even RQFT practitioners) only deal with RQFT in the context of
specific scientific problems, reinforcing the perception of the
theory as something rather esoteric and far removed from everyday
phenomena.

In order to understand points (a) and (b) in more detail, it is
instructive to review the way this outlook is imparted to Physics
students. One starts by studying non-relativistic and then
relativistic Classical Mechanics (CM), followed by classical wave
motion and Classical Electrodynamics (CED).  One can even learn
General Relativity (GR), still without any mention of quantum
physics. One is then usually given the impression that almost all
macroscopic physical phenomena can be described very successfully
within this classical framework, except for isolated shortcomings
such as its inability to account for a few obscure experimental
facts involving black-body radiation (curiously, the reason the
Rayleigh-Jeans prediction truly deserves to be called a
catastrophe is not sufficiently emphasized-the classical theory
predicts that not only black-bodies, but all substances at
non-zero temperature continually emit an infinite amount of
energy!), photoelectric and Compton effects, and the discrete
emission and absorption spectra of elements. In order to solve
these problems supposedly limited to atomic or sub-atomic scales,
where classical physics fails, QM is invoked. The removal of finer
discrepancies between experiment and the predictions of
non-relativistic QM require corrections based first on
relativistic QM, and only then finally on RQFT. These are
accompanied by increasing degrees of conceptual and computational
difficulty. Given this order of exposition, one can easily come
away with the idea that RQFT is a formidably complex theory with
little direct relevance to our everyday experience, which is
carelessly claimed to be adequately described by Classical
Physics. This view is bolstered by formal demonstrations that the
classical laws of Physics are recovered in the limits of quantum
numbers tending to infinity and Planck's constant $h$ going to
zero.  In the course of such arguments, one is sometimes inclined
to forget that Planck's constant, while small on a macroscopic
scale, is not exactly zero, and the macroscopic world is actually
shaped by that fact.

The picture summarized in the last paragraph, while perhaps free
from errors of commission, seriously suffers from a major error of
omission: It completely overlooks the fact that direct
manifestations of RQFT are at the root of our everyday
experiences. An independent inaccuracy stems from the popular but
false assertion according to which Quantum Mechanics is relevant
only for physical events on a microscopic (atomic or subatomic)
scale, without referring to the phase relationships in the
problem. A more accurate classification of physical phenomena must
take into account not only whether they occur on a macroscopic or
a microscopic scale, but also whether there is coherence or
decoherence.  This gives rise to four combinations which we may
label as (i) macroscopic-coherent, (ii) microscopic-coherent,
(iii) microscopic-incoherent and (iv) macroscopic-incoherent.
Lasers, Bose-Einstein-condensation, superconductivity and
superfluidity are examples of the first, double-slit experiments
with electron sources examples of the second, and any Quantum
Mechanical problem in the microscopic domain, described
successfully with random phases, is an example of the third.
Finally, everyday macroscopic phenomena involving the
participation of something on the order of Avogadro's number of
particles with uncorrelated phases belong to the last category. It
is often claimed that classical physics suffices to describe this
domain. We will argue later that quite the opposite is the case:
Attempting to imagine a universe truly and strictly based upon
classical physics leads to a world far stranger than the one we
live in.  Our familiar environment and the way we think about it
are shaped by macroscopic manifestations of RQFT; it is only the
phases that are washed out by decoherence.

\noindent {\bf 2.  Some childlike questions concerning identity,
similarity, integers, common names and verbs :}

One possible way to start removing the lifelong accumulation of
curiosity-deadening layers of familiarity with the everyday world
is to ask questions similar to those sometimes raised by children,
and then to address in turn the new questions raised by the first
round of answers. Let us start with the following :

Q1: Why are identical twins identical?

Q2: Why do we encounter categories of objects that we can classify
in groups such as spoons, cats and clouds?

Q3: What is the origin of the concept of natural numbers?

Q4: How does a glass of water get filled, and, more generally, why
do objects in the condensed state occupy well-defined volumes?

These innocent-sounding questions are actually intimately linked
to timeless philosophical and scientific issues.  For example, the
question of identity mentioned in Q1 has been examined by Leibniz
\cite{LEIB}, while Plato tried to answer Q2 in terms of
``Universals". This roughly means that spoons, cats and clouds are
only imperfect copies of an Ideal Spoon, an Ideal Cat and an Ideal
Cloud existing in a ``higher" and physically inaccessible realm,
of whose existence we nevertheless have an imperfect awareness
\cite{PLAT}. After Plato, the question of Universals continued to
be a central concern in medieval scholastic philosophy and, in
modified forms, is debated even today among philosophers. Whatever
one thinks of Plato's explanation, it is a fact that we not only
recognize such categories, but even incorporate them into our
conscious or unconscious mental processes, and indeed into our
language, with which we formulate philosophical problems.

Turning to Q3, one may wonder how the concept of natural numbers
could have arisen without the existence of the above categories.
In fact, Frege \cite{FREG} tried to define, say, the natural
number 3, as the property common to sets of three spoons, three
cats and so on; it is difficult to imagine how such a notion could
have emerged in a universe consisting of amorphous objects, each
unlike any other. The fact that this definition involves
set-theoretic paradoxes that had to be resolved by Russell and
Whitehead \cite{RUSSELL} is an irrelevant complication for the
purposes of our specific argument.

Having noted that the concept of number is linked to the existence
of common (as opposed to proper) names, and traced the origin of
at least some of the common names to the very strong physical
similarities of the objects they represent, we can carry out a
parallel analysis about verbs, with parallel conclusions. For
example, we can recognize the action of eating in many life-forms
and classify all such activity under the verb ``to eat".
Essentially, since the agents of the actions fall into categories
labelled by common names, so do the actions.

Finally, Q4 is obviously related to fundamental notions of
``full'' and ``empty", which not only are basic to our everyday
thinking, but lead to questions concerning the definitions of
``Matter" versus  ``Nothingness"; let us call this question Q5.
Descartes' claim that Nature abhors a vacuum is only one example
in a series of changing pictures of empty space conceived by
philosophers and scientists. Physicists have gone from Newton's
vacuum to Maxwell's luminiferous Aether \cite{WHITTAKER}
supporting electromagnetic waves, then back to a plain empty
vacuum, and finally to modern versions of Aether-like media
constituted of the Dirac sea, vacuum expectation values of Higgs
fields, and superpositions of topologically inequivalent vacua. We
can see RQFT themes already appearing in this last sentence; it
will soon become clear that this is no accident.

Finally, it is certainly not far-fetched to assume that many of
our geometrical ideas originate from the study of shapes of
objects formed by matter in the condensed state; we will see in
detail later on that many essential properties of the condensed
state are direct consequences of the RQFT of electrons.

A disclaimer is perhaps in order at this point: our arguments do
not preclude the possibility, or indeed the certainty, that there
is a deeper physical theory which yields RQFT as an approximation
at lengths much bigger than the Planck length
$\sqrt{\frac{hG}{c^3}}$ ; in fact, the absence of a quantum-level
explanation of gravitation points to the need for such a theory.
However, such a theory, if and when found, must also necessarily
reproduce the successful features of RQFT pertinent to the main
line of thought pursued here.

\noindent {\bf 3. Some answers and more questions:}

Let us start with Q1, which could not have been handled prior to
the understanding of DNA's role in heredity.  The modern and
conclusive answer is of course the identity of twins' DNAs. This
of course also provides a less fantastic explanation of the
obvious similarities between cats than Plato's postulation of an
Ideal Cat in the World of Forms: Ignoring minor variations
corresponding to different colors and sizes and so on, cat DNAs
are very alike in containing instructions for one cat nose, one
pair of cat eyes and one pair of cat ears per cat.  To turn to
question Q2 of the previous section, clearly, the ``Platonic
spoon" can be similarly explained: Since all human mouths and
hands are similar in size and function, and since only a finite
number of chemical substances are suitable for being formed into
spoons, an ``induced universality" operates even in the case of
such man-made objects. Even a member of a category such as
``rocks", consisting of relatively amorphous objects, is
classifiable (a) because rocks are obtained from a limited number
of elements via a limited number of geological processes, (b) our
perceptions of their salient characteristics such as hardness are
bound to be similar, given the similarity in human bodies and, in
particular, in human sensory equipment.

These remarks immediately raise other questions such as why DNA is
so stable and how it happens that the DNAs of identical twins are
practically identical at the molecular level.  To our knowledge,
such questions were first addressed by Schr\"{o}dinger in his very
influential book "What is Life?" \cite{SCHRO}.  His answer is
based essentially on the discreteness of molecular energy levels
for quantum mechanical bound systems and the differences $\Delta
E_{ij}$ between energy levels $E_{i}$ and $E_{j}$ of a DNA
molecule typically being considerably bigger than the thermal
energy $kT$ per degree of freedom.  Very schematically, we might
say that if the level $E_{i}$ corresponds to a cat with one nose
and the higher level $E_{j}$ to one with two noses, random thermal
and most other environmental influences will very rarely be able
to cause a transition resulting in a two-nosed cat. There are more
modern and detailed explanations of the stability of DNA that are
based on its being copied by enzymes from a template and other
enzymes correcting errors, etc., but for our purposes these may be
regarded as elaborations rather than refutations of
Schr\"{o}dinger's fundamental insight.

While Schr\"{o}dinger's argument accounts for the stability of a
single lone DNA molecule in the entire universe, accounting for
the existence of other identical DNA molecules obviously requires
that we now explain why all the atoms of a given chemical element
are perfectly identical. This is essentially impossible in
Classical Mechanics, and even Quantum Mechanics by itself provides
only a partial explanation: For example, given a single proton and
an electron, Quantum Mechanics predicts, via the Schr\"{o}dinger
equation, a definite set of discrete energy levels for a single
Hydrogen atom, but does not explain why there are other electrons
and protons with exactly the same properties, combining to form
other Hydrogen atoms.  It hardly needs to be emphasized that the
existence of categories of non-living things such as spoons and
rocks that we mentioned earlier also rely on all the atoms of a
given element being identical with all the others; hence we must
next see what lies behind the identity of the constituents of
atoms.

Let us introduce the argument by focusing upon electrons in
particular. It is well known that Wheeler attributed the identity
of electrons to there being only one electron in the universe
\cite{FEYNMAN}. This is of course based on Dirac's Hole Theory,
with Wheeler's additional observation that a positron travelling
from the past to the future can be thought of as a negative energy
electron travelling from the future to the past. The intersection
points of a constant-time hyperplane with a single electron
world-line going back and forth in time (with photons attached to
the vertices) then represent simultaneously present electrons and
positrons. Although this is a fascinating, and even useful idea
(having led as a practical consequence to the Feynman propagator),
it cannot even describe all the physical processes involving only
electrons, positrons and photons.  As an example of a process
requiring a second electron unrelated to the first, consider
adding to the zig-zagging electron picture a spacetime diagram of
Delbr\"{u}ck scattering at lowest order. The diagram consists of
an electron going around a square, with a photon at each corner.
The inability of Wheeler's scheme to account for such processes is
of course a manifestation of the fact that combining Special
Relativity with Quantum Mechanics inevitably leads to a
many-particle theory. RQFT is intrinsically suited to describing
such many-particle systems, and we thus need not dwell any longer
in Wheeler's halfway house on the way to QED, attractive though it
is. Nevertheless, given our basic theme of relating fundamental
philosophical concepts to RQFT, we should take note of the
completely radical ontological switch implied by Hole Theory: What
is normally regarded as empty space is identified with a negative
energy electron sea of infinite negative charge and energy, while
a hole in this sea appears to be a physical particle of positive
energy and charge. Also, although we have not yet fully gone over
to RQFT, we should take note of two fundamental ingredients,
namely, Special Relativity and the Pauli Exclusion Principle
(PEP), which already play an essential role in the above
arguments. Hole theory is based on taking the negative sign in the
relativistic expression $ E = \pm\sqrt{p^2 c^2 + m^2 c^4}$
seriously.  The existence of ordinary positive energy electrons
then depends on all the negative energy levels already being full,
and the filling process in turn is made possible by the PEP,
without which the negative levels would turn into a bottomless pit
(we should note in passing that this ingenious argument of Dirac
implicitly assumes the existence of a minimum energy level, which,
inexplicably, is never mentioned).

Finally, the PEP is an essential part of the answer to Q4 of the
previous section.  Filling a glass of water, surely one of our
most ordinary everyday experiences, is a purely Quantum Mechanical
phenomenon: The electrons of one water molecule are kept away from
those of the other by the PEP. The molecules themselves are mostly
"empty" in the sense that all but a few ten-thousandths of a
molecule's mass is in the nucleus, which typically occupies only
about one million billionth of the volume of the molecule (the
volume of the Sun relative to the volume defined by the Earth's
orbital radius is vastly bigger!).  The reason an individual atom
does not collapse down to the size of its nucleus again involves
both the PEP and the Heisenberg Uncertainty Principle (HUP).  Thus
our very basic notions of ``full" and ``empty" are seen to be
based entirely on two fundamental Quantum Mechanical rules, the
HUP and the PEP. When we speak of filling the electronic levels in
an atom, we are not just using an everyday idea as a metaphor for
a phenomenon involving subatomic particles; it is the microscopic
phenomenon that is responsible for our having the adjective
``full" in our vocabulary. It is hard to imagine the idea of
fullness ever arising in a universe consisting of bosons, which
are not subject to the PEP.

A more dramatic example is provided by the numerous suicides that
take place near Bo\u{g}azi\c{c}i University, where I work, every
year. The majority of the people jumping off the nearby bridge
linking Asia and Europe die from the impact of their bodies
hitting the water. Given that both the region within their bodies
and the water they fail to displace sufficiently rapidly consists
essentially of empty space, the deaths are directly attributable
to two basic Quantum Mechanical principles, the HUP and the PEP
again. We will later see that both find their most natural
expressions in RQFT.

\noindent {\bf 4.  RQFT:}

In order to explain the identity of electrons in terms of RQFT, we
observe first that the main dynamical entity in a Field Theory is
the field itself \cite{INFELD}. This is a very different viewpoint
than the ``particle picture", which, even after relativistic and
Quantum Mechanical refinements and modifications, still uses the
idea of a pointlike particle in the presence of external forces as
its basic ingredient.  In contrast, whether classical or quantum,
relativistic or non-relativistic, a typical field is thought of as
a dynamical system pervading all of space for all times.   The
further qualifying words ``relativistic" and ``quantum" have very
specific meanings: By ``relativistic" here we mean a theory in
which the fields and states transform as well-defined
representations of the Poincar\'{e} group.  We thus limit the
discussion to flat Minkowski space and neglect effects of
space-time curvature.  This description holds to an excellent
approximation in our everyday physical environment.  The quantum
nature of the theory can be represented in a number of different
and more or less equivalent ways such as canonical or
path-integral quantization. We will adopt the canonical approach
based on equal-time commutators or anticommutators, as it
expresses most directly the features we intend to emphasize. Dirac
\cite{DIRAC} aptly calls these (anti)commutators ``the quantum
conditions" since it is through them that quantum characteristics
are imparted to a theory initially formulated in classical terms.
The HUP is in fact a consequence of the quantum conditions in
particle Quantum Mechanics; turning this around, we might say that
field quantization via the canonical method amounts to extending
the HUP from particle Quantum Mechanics to Field Theory.

The building of a RQFT within the above framework makes use of a
number of fundamental theorems and mathematical techniques such as
Wigner's method of obtaining and classifying the representations
of the Poincar\'{e} group, Pauli's Spin-Statistics
theorem\cite{PAULI}, derivation of the Bargmann-Wigner
\cite{BARG-WIG} equations for fields of a given spin and Noether's
theorem.   To the extent it is possible, we will try to rely on
qualitative arguments in tracing the questions of section 2 to
this formidable-sounding collection of mathematical physics
results. This will also be helpful in allowing us to see direct
consequences of the formalism in the everyday world without being
lost in sophisticated and abstract mathematics.

Let us start with the Poincar\'{e} group. As emphasized by Klein
and Weyl\cite{SYMMETRY}, a group is a collection of operations
leaving a certain ``object" unchanged.  This amounts to
classifying the symmetries of the object. When the ``object" in
question is the laws of Physics in a space-time with negligible
gravitation-induced curvature, the symmetries can be classified as
follows: (i) No point in four-dimensional space-time is
privileged, hence one can shift or translate the origin of
space-time arbitrarily in four directions. Noether's theorem then
implies there are four associated conserved quantities, namely the
three components of space momentum and the energy. These four
quantities naturally constitute the components of a 4-vector
$P_{\mu}, \mu = 0,1,2,3$. (ii) No direction is special in space;
leading to three conserved quantities $J_{i},i=1,2,3$. (iii) There
is no special inertial frame; the same laws of Physics hold in
inertial frames moving with constant speed in any one of the three
independent directions.

As we suggested above, it is possible to get a non-mathematical
insight into Noether's theorem relating symmetries to conserved
quantities; in fact the argument we will present, due to John
Philoponus \cite{PHIL}, dates back to the 6th century AD! Consider
a single particle moving in a completely homogeneous space.  It
cannot come to a stop or change its velocity because this would
have to happen at some particular point, but all points being
equal, it is impossible to choose one. Hence the particle has no
choice but to move at constant velocity or, in other words, to
conserve its linear momentum, which Philoponus called ``impetus".
It is easy to extend the argument to a rotating object in an
isotropic space and conclude that it cannot come to a stop at any
particular angle since there is no special angle; hence its
angular momentum is conserved.

It is well-known that (ii) and (iii) amount to covariance of the
laws of physics under rotations in a four-dimensional space with a
metric that is not positive-definite. The squared length of a
4-vector defined via this metric must then be an important
invariant independent of the orientation or the velocity of the
frame. Indeed, for the 4-vector $P_\mu$this is the squared mass
$m^2$ of the particle, and it is one of the two invariant labels
used in specifying the representation.  The other label is the
squared length of another 4-vector called the Pauli-Lubanski
vector.  It then follows from the algebra of the group that this
squared length takes on values $s(s+1)$ and that in contrast to
$m^2$, which assumes continuous values, $s$ can only be zero, or a
positive integer, or half a positive odd integer. The unitary
representation of the Poincar\'{e} group for a particle of mass
$m$ and spin $s$ provides its relativistic quantum mechanical wave
function. The equation of motion the wave function must obey also
comes with the representation; it is the Bargmann-Wigner equation
for that spin and mass. The procedure of second quantization then
naturally promotes the wave functions to quantized field
operators, and in a sense demotes the particles to quanta created
or destroyed by these operators.  Pauli's spin-statistics theorem,
based on a set of very general requirements such as the existence
of a lowest energy vacuum state, the positivity of energy and
probability, microcausality, and the invariance of the laws of
Physics under the Poincar\'{e} group, leads to the result that the
only acceptable quantum conditions for field operators of
integer-spin particles are commutation relations, while those
corresponding to half-integer spin must obey anticommutation
relations. The standard terms for the two families of particles
are bosons and fermions, respectively.  The PEP, or the
impossibility of putting two electrons into the same state, is now
seen to be the result of the anticommutation relation between
electron creation operators: to place two fermions in the same
state, the same creation operator has to be applied twice.  The
result must vanish, since the operator anticommutes with itself.

We have thus seen that the symmetries of space-time are reflected
in the fields which are representations of the symmetry groups; a
quantum mechanical recipe called quantization then turns these
fields into operators capable of creating and destroying quanta
(or particles, in more common parlance) at all space-time points.
Actually, the framework we have described only suffices to
describe ``Free fields" which do not interact with each other. In
order to incorporate interactions, one has to resort to another
kind of symmetry called gauge symmetry, which operates in an
``internal" space attached to each point of space-time.  While the
identity of masses and spins of, say, electrons can be attributed
to space-time symmetries, the identities of additional quantum
numbers such as charge, isospin and ``color" can only be explained
in terms of the representations of these gauge groups.

%%%%%%%%%%%%%%%%%%%%%%%%%%%%%%%%%%%%%%%%%%%%%%%%%%%%%%%%%%%%%%%%%%%
\noindent {\bf 5. The answers:}

We are now in a position to relate the questions of sections 2 and
3 to RQFT.  Electrons are identical because they are created by
the same electron quantum field. Since this field embodies
Poincar\'{e} invariance, it operates in the same way at all
space-time points and in different Lorentz frames; thus the
electrons created by it must always have the same spin, mass and
charge. The same obviously holds for other fermions such as the
quarks, which are first bound together by bosonic fields called
gluons to form protons and neutrons. These nucleons then combine
to form atomic nuclei through Van der Waals-like residual gluonic
forces. At suitable densities and temperatures, such as $300,000$
years after the Big Bang, hydrogen and helium nuclei can bind
electrons by photon fields and form atoms. Heavier atoms result
from processes of stellar evolution and collapse.  Atoms
chemically bond with other atoms, producing ever more complex
molecules and, finally, intelligent life which tries to understand
its environment. Every step in the dynamical processes outlined
above involves not only the forces, but also the constraints
imposed by the commutation and/or anticommutation relations of the
fields. The stability of matter, in particular, is in large
measure due to the PEP.  Free neutrons undergo $\beta$-decay with
a lifetime of about ten minutes, but live practically forever in
stable nuclei because the proton that would result from the decay
cannot move into energetically available but already occupied
proton states. The arrangement of electrons in atoms and the
formation of molecular bonds, both ionic and covalent, is dictated
by the PEP. The near-incompressibility of most condensed matter,
the stability of white dwarf and neutron stars, are among its
macroscopic consequences.  These qualitative statements are
supported by rigorous arguments and computations \cite{LIEB}

To recapitulate:  in section 2, we raised some questions about the
origins of common names (or, equivalently, the Platonic concept of
Universals) and verbs (Q1 and Q2)), the idea of whole numbers
(Q3), and our perception of regions of space as ``full" and
``empty" (Q4). Q5, which was related to Q4, concerned the relation
between ``being" and ``nothingness", the former in the specific
example of, an electron, and the latter, the vacuum.  In section
3, we gave partial answers based on Non-Relativistic and then
Relativistic Quantum Mechanics, supplanted by the PEP, which had
to be introduced as an additional, empirically-based input.  The
answers were partial in the sense that while, for example, the
identity of the energy levels in two hydrogen atoms could be
explained, the identity of all electrons, or all protons, had to
be assumed as a starting fact. In relation to question Q5, we gave
an example of a theory where the vacuum is represented by a full
medium, while holes in the medium are detected by experimenters as
positively charged versions of the electrons. Finally, in section
4, we saw that the identity of fundamental particles is a direct
consequence of the symmetries of space-time and the quantum
conditions in RQFT. Furthermore, the PEP was revealed to be
nothing but the quantum conditions for fermion fields; hence we
can now quite accurately pinpoint the fundamental cause of death
when people jump off the Bosphorus bridge: The killer is the
equal-time anticommutation relation of the quantized electron
field. We hope to have convinced the reader that RQFT, far from
being a theory limited to explaining esoteric subatomic phenomena,
shapes both the world we directly experience and our most
fundamental concepts about it.

So far we have limited ourselves to a picture in which a
disembodied intelligence acquires ideas about the external world
as a result of its experiences.  Let us now make assumption, by no
means universally accepted by philosophers, that all mental events
are rooted in biochemical brain events, and delve into the actual
workings of the human brain on the basis of that assumption. While
brain science is in its infancy, one of the known facts is that
electrical signals are sent among nerve cells along channels
through variations in the concentrations of $Na^+$, $K^+$ and
$Cl^-$ ions inside and outside the channels \cite{Penrose}. Given
that the code for constructing the brain is encoded in the DNA
which we argued owes it stability and repeatable features
ultimately to RQFT, and that the same holds for the ions used in
the signalling, we see that human minds, great or not, think alike
not just because of the similarities in the data they receive from
a RQFT-shaped environment, but because they themselves are also
ultimately shaped by RQFT.

\noindent {\bf 6.  A truly classical universe?}

Let us now try to imagine a Universe in which Planck's constant
$h$ is not merely small compared to macroscopic actions or angular
momenta, but truly and strictly zero. The laws of classical
Physics that can be derived formally from their quantum mechanical
versions by focusing on expectation values and letting $h$ go to
zero (even General Relativity, whose quantum origins are not yet
completely understood, has a tentative microscopic version in
String theory) will now, for the sake of argument, be regarded as
being ``exact" and ``fundamental".  These laws consist of (i)
Einstein's field equations, with the stress-energy tensors of the
electromagnetic field and of massive particles on the right-hand
side, (ii) geodesic equations for the motion of particles
(ignoring the question of how such particles may come into being
within classical physics) in gravitational and electromagnetic
field backgrounds and (iii) Maxwell's equations in curved space
for electromagnetic fields, with charge-current sources again on
the right-hand side.  Such a system of equations has well-known
problems such as runaway solutions and infinite self-energies for
particles, so in a sense it is not entirely logical to pursue this
line of thought. Nevertheless, let us temporarily ignore these
objections in order to sketch the qualitative aspects of the
physical environment the equations predict. A basic feature of
such a universe must be the loss of energy of accelerating matter
through electromagnetic or gravitational radiation.  The
slowed-down masses with or without charges will then tend to clump
together under attractive gravitational and electromagnetic
forces. Unlike in quantum physics, there will be no
counterbalancing effects such as the HUP or the PEP, and collapses
are bound to occur even if they may be delayed by initial
velocities and angular momenta. The only imaginable result is a
collection of Black Holes, each one of which has angular momentum,
mass and charge unlike any other.  There being no Planck's
constant, such Black Holes cannot evaporate via Hawking radiation.
The initial conditions will then determine whether the Black Holes
will eventually diverge from each other in an open universe or all
will collapse into a single hole.   Provided it is physically
possible at all, this is the likely actual picture of Nature
envisaged by ``Classical Physics", but it is clearly does not
resemble the world we are a part of.

\noindent {\bf Acknowledgements}

I am grateful to Ali Alpar and Erdal In\"{o}n\"{u} for helpful
discussions and encouragement.

\end{document}